\documentclass[%
 aip,
 float,
 amsmath,amssymb,
 reprint,%
]{revtex4-1}

\usepackage{graphicx}% Include figure files
\usepackage{dcolumn}% Align table columns on decimal point
\usepackage{bm}% bold math
\usepackage{hyperref}
\usepackage{upgreek}

\usepackage[utf8]{inputenc}
\usepackage[T1]{fontenc}
\usepackage{mathptmx}
\usepackage{etoolbox}
\usepackage{siunitx}
\usepackage{comment}
\usepackage[caption=false]{subfig} % <=== "caption=false" added
\usepackage{float}
\usepackage{setspace}
\makeatletter
\def\@email#1#2{%
 \endgroup
 \patchcmd{\titleblock@produce}
  {\frontmatter@RRAPformat}
  {\frontmatter@RRAPformat{\produce@RRAP{*#1\href{mailto:#2}{#2}}}\frontmatter@RRAPformat}
  {}{}
}%
\makeatother
\begin{document}

\preprint{AIP/123-QED}

% Force line breaks with \\
\title{A highly stable and fully tunable open microcavity platform at cryogenic temperatures}

\author{Maximilian Pallmann$^\dag$}
\affiliation{ 
Physikalisches Institut, Karlsruhe Institute of Technology (KIT), Wolfgang-Gaede Str. 1, 76131 Karlsruhe, Germany}

\author{Timon Eichhorn$^\dag$}
\affiliation{ 
Physikalisches Institut, Karlsruhe Institute of Technology (KIT), Wolfgang-Gaede Str. 1, 76131 Karlsruhe, Germany}

\author{Julia Benedikter}
\affiliation{Faculty of Physics, Ludwig-Maximilians-University (LMU), Schellingstr. 4, 80799 Munich, Germany}

\author{Bernardo Casabone}
\affiliation{ICFO-Institut de Ciencies Fotoniques, The Barcelona Institute of Science and Technology, 08860 Castelldefels, Barcelona, Spain}

\author{Thomas Hümmer}
\affiliation{Faculty of Physics, Ludwig-Maximilians-University (LMU), Schellingstr. 4, 80799 Munich, Germany}
\affiliation{Qlibri GmbH, Maistr. 67, 80337 Munich, Germany}

\author{David Hunger}
\affiliation{ 
Physikalisches Institut, Karlsruhe Institute of Technology (KIT), Wolfgang-Gaede Str. 1, 76131 Karlsruhe, Germany}
\affiliation{Institute for Quantum Materials and Technologies (IQMT), Karlsruhe Institute of Technology (KIT), Herrmann-von-Helmholtz Platz 1, 76344 Eggenstein-Leopoldshafen, Germany}
\email{david.hunger@kit.edu}

\date{\today}% It is always \today, today,

\begin{abstract}
$\dag$ These authors contributed equally. \newline
\newline
\noindent
Open-access microcavities are a powerful tool to enhance light-matter interactions for solid-state quantum and nano systems and are key to advance applications in quantum technologies. For this purpose, the cavities should simultaneously meet two conflicting requirements - full tunability to cope with spatial and spectral inhomogeneities of a material, and highest stability under operation in a cryogenic environment to maintain resonance conditions. To tackle this challenge, we have developed a fully-tunable, open-access, fiber-based Fabry-Pérot microcavity platform which can be operated also under increased noise levels in a closed-cycle cryostat. It comprises custom-designed monolithic micro- and nanopositioning elements with up to mm-scale travel range that achieve a passive cavity length stability at low temperature of only \SI{15}{pm} rms in a closed-cycle cryostat, and \SI{5}{pm} in a more quiet flow cryostat. This can be further improved by active stabilization, and even higher stability is obtained under direct mechanical contact between the cavity mirrors, yielding \SI{0.8}{pm} rms during the quiet phase of the closed-cycle cryo cooler. The platform provides operation of cryogenic cavities with high finesse and small mode volume for strong enhancement of light-matter interactions, opening up novel possibilities for experiments with a great variety of quantum and nano materials.

\end{abstract}

\maketitle

\section{Introduction}
\label{sec:introduction}

\begin{comment}
General intro structure (1-2 sentences per point):
\begin{enumerate}
    \item (Why quantum technologies?)
    \item (Why quantum optics?)
    \item Different samples for open-access FPIs: nanoparticles, TMDs, color centers, REI, QDs
    \item Why cavities, cavity QED?
    \item Why tunable open-access FP microcavities?
    \item Why cryogenics, closed-cycle cryostat?
    \item Numbers on required mechanical stability
    \item What is the state-of-the-art (status quo)?
    \item "In this work, we..." summarize our achievements
\end{enumerate}
\end{comment}
%add more citations for nanocavities and hBnhttps://www.overleaf.com/project/6226120c77c81f75c89f42bd
Open-access optical microcavities are a versatile platform to enable the enhancement of light-matter interactions for quantum and nano systems with a broad range of applications such as efficient single photon sources \cite{tomm_bright_2021,johnson_diamond_2017,salz_cryogenic_2020,hausler_tunable_2021,vogl_compact_2019,benedikter_cavity-enhanced_2017,albrecht_coupling_2013} and spin-photon interfaces \cite{janitz_fabry-perot_2015,riedel_deterministic_2017,ruf_resonant_2021}, single emitter strong coupling \cite{najer_gated_2019,wang_turning_2019}, exciton polaritons \cite{li_tunable_2019,dufferwiel_exciton-polaritons_2015,flatten_room-temperature_2016,gebhardt_polariton_2019}, and cavity optomechanics \cite{aspelmeyer_cavity_2014,a_d_kashkanova_superfluid_2016,rochau_dynamical_2021}. In order to achieve good coherence properties of the material, most of them have to be cooled down to cryogenic temperatures. For the example of solid state quantum emitters, stringent requirements on the coherence, particular transition frequency, local spin environment etc. typically require a careful selection of suitable emitters from a large set of candidates. A cavity platform integrated in a cryostat that is capable of finding \textit{in situ} suitable emitters on such spatially inhomogeneous samples and addressing optical transitions across a large spectral range is thus highly desirable. 

%Fully-tunable open-access fiber-based Fabry-Pérot microcavities (FFPC) \cite{hunger_fiber_2010}.

%A variety of different kinds of samples is currently being investigated, and examples include individual quantum emitters such as color centers in diamond \cite{benedikter_cavity-enhanced_2017,ruf_resonant_2021,hoy_jensen_cavity-enhanced_2020}, rare earth ions \cite{casabone_cavity-enhanced_2020,merkel_coherent_2020}, semiconductor quantum dots \cite{najer_gated_2019,}, transition metal dichalcogenides \cite{} two-dimensional 
%such as rare-earth ion (REI) doped nanoparticles \cite{casabone_cavity-enhanced_2018}, membranes \cite{merkel_coherent_2020} or thin-films \cite{harada_controlling_2022}, color centers in diamond membranes \cite{benedikter_cavity-enhanced_2017,ruf_resonant_2021,hoy_jensen_cavity-enhanced_2020}, transition-metal-dechalcogenides (TMD) \cite{vadia_open-cavity_2021} or semiconductor quantum dots \cite{najer_gated_2019}. A cavity platform that is capable of finding suitable emitters on such spatially inhomogeneous samples and addressing optical transitions ranging over the visible to near-infrared spectrum are fully-tunable open-access fiber-based Fabry-Pérot microcavities (FFPC) \cite{hunger_fiber_2010}.\\

\begin{figure}
    \centering
    \includegraphics[width=0.34\textwidth]{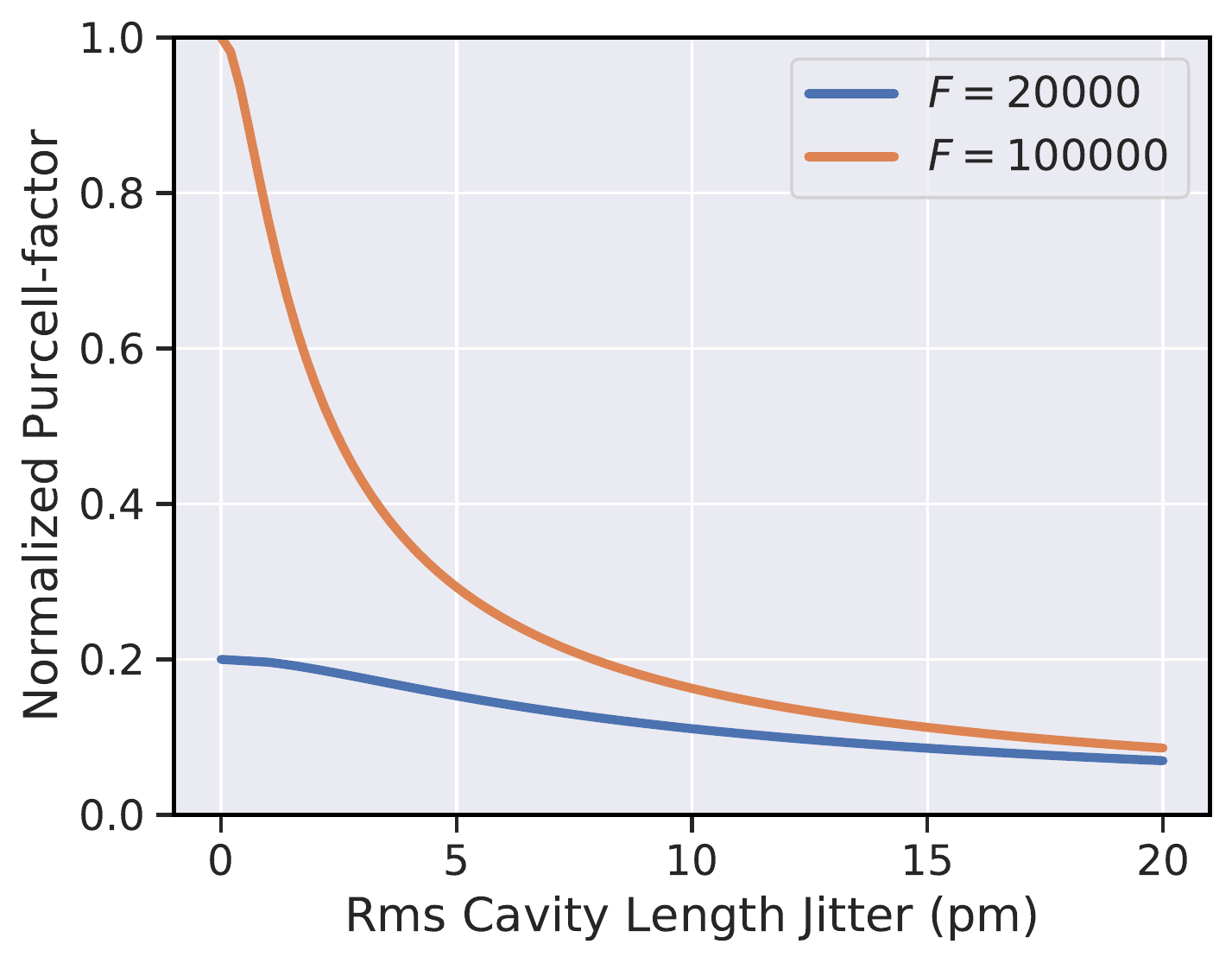}
    \caption[]{Dependence of the averaged Purcell-factor for a cavity of 20,000 Finesse (blue) and 100,000 (orange) on the rms cavity length jitter. Both curves are normalized to the maximum Purcell-factor for $F= 100,000$.}
    \label{fig:Purcell_vs_RMS}
\end{figure}

Using a closed-cycle cryostat provides advantages in terms of usability as well as scalability, but poses a challenge due to the increased mechanical noise level compared to wet cryostats. Although current commercial solutions offer mechanical vibrations of the experimental cold plate down to a few \SI{}{nm} root mean square (rms), this still remains a severe challenge for operating a high-finesse microcavity. For example, maintaining resonance conditions for a cavity with a finesse of $F=20,000$ at a wavelength of \SI{580}{nm} requires a cavity length stability of less than the spatial linewidth  $\Delta L=\lambda/2F=15$ pm, i.e. three orders of magnitude lower than the vibrations of the cold plate. The figure of merit for the light-matter interaction between the cavity photons and the quantum emitter is the Purcell-factor $C\propto F/w_0^2$ with $w_0$ the cavity mode waist at the emitter. Assuming a Gaussian broadening of the cavity linewidth due to mechanical vibrations, the time-averaged Purcell-factor of the system $C$ is decreased with respect to the resonant Purcell factor $C_0$ according to \cite{fontana_mechanically_2021}
\begin{align}
    \frac{C(\Delta L, \sigma)}{C_0} = \sqrt{\frac{\pi}{8}} \frac{\Delta L}{\sigma} e^{\frac{\Delta L^2}{8\sigma^2}} \left (1-\mathrm{erf}\left (\frac{\Delta L}{2\sqrt{2}\sigma}\right) \right),
\end{align}
where $\sigma$ denotes the rms cavity length jitter. This effective Purcell-factor is plotted in Fig.~\ref{fig:Purcell_vs_RMS} for a cavity Finesse of \SI{20,000}{} and \SI{100,000}{}, respectively, normalized to the latter. From this one can see that a length stability corresponding to one cavity linewidth $\Delta L$ leads to a reduction to \SI{40}{\percent} of the resonant Purcell factor. The aim is therefore to minimize the length jitter to the level of a few \SI{}{pm}.\\

In the past years, significant progress has been made to reach such high mechanical stability by using passive vibration isolation techniques as well as active feedback  \cite{casabone_dynamic_2021,fontana_mechanically_2021,vadia_open-cavity_2021,ruf_resonant_2021,ruelle_tunable_2022,najer_gated_2019,merkel_coherent_2020}. However, for platforms operated in closed-cycle cryostats, the best stability reached so far is about \SI{14}{pm} rms during the quiet phase of the cryostat cycle and using active stabilization \cite{fontana_mechanically_2021}.\\

In this work, we present two variants of custom-designed, fully-tunable open-access Fabry-Pérot microcavities that are designed for the operation in a flow cryostat and a closed-cycle cryostat, respectively. At room temperature and when the cryostat is switched off, we reach a stability of \SI{1}{pm} rms under active stabilization. At cryogenic temperatures and operated in the flow (closed-cycle) cryostat, we achieve a passive stability of \SI{5 (15)}{pm} rms. This makes it possible to perform experiments without the need for active stabilization, which typically requires a separate locking laser at a different wavelength that limits the usable cavity lengths. Active stabilization can further improve this, and we achieve a stability of \SI{2.5 (15)}{pm} rms in the flow (closed-cycle) cryostat. Operating the cavity in a contact mode by bringing the fiber mirror into mechanical contact with the planar opponent, we achieve a stability of \SI{0.7}{pm} rms in the flow cryostat and \SI{0.8}{pm} rms during the quiet phase of the closed-cycle cryostat cycle. %We can prove the performance reproducibility of our mechanical design by reaching also very high passive stabilities of about \SI{7}{pm} rms inside the flow cryostat without bringing both mirrors into contact. \\

\begin{figure*}
    \centering
    \includegraphics[scale=0.35]{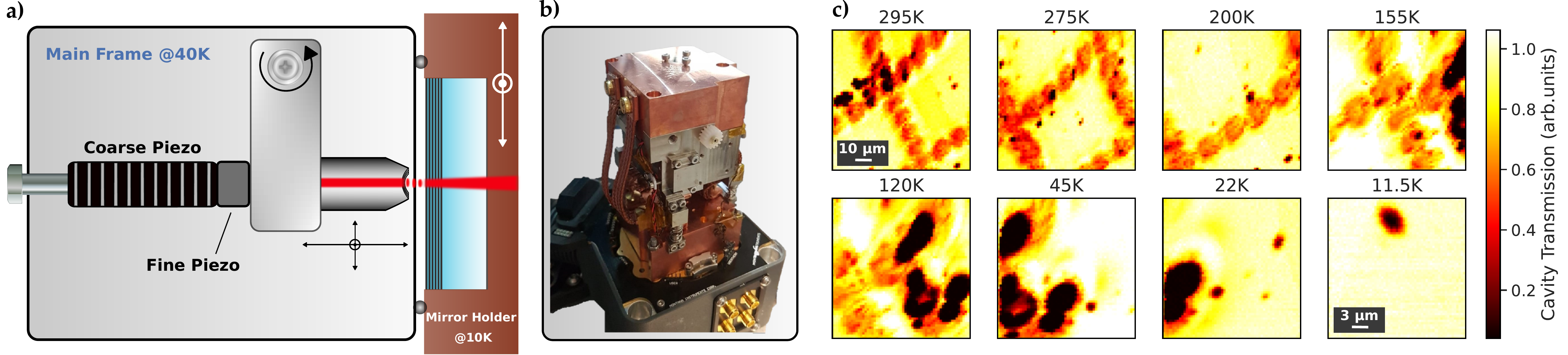}
    \caption{\textbf{(a)} Schematic drawing of the cavity setup: The planar DBR-mirror (blue) and the fiber mirror can be moved laterally to select a region of interest on the sample. The mirror separation is controlled by two combined piezo stacks, referred to as coarse and fine piezo, that push the lever holding the fiber.
    \textbf{(b)} Photograph of the closed-cycle version of the cavity stage. \textbf{(c)} Cavity transmission scans of a marker grid during a cooldown of the closed-cycle cryostat. When reaching the final temperature, the fiber position has left the grid, but a scattering particle on the planar mirror reveals the cavity point spread function.}
    \label{fig:setup_and_cd_scans}
\end{figure*}

\begin{comment}
Motivation: why using cavities?: Strong-coupling, Purcell-enhancement, extinction \textrightarrow{} why using fiber FPI?: high Cooperativity through small mode volume + high Q, possible for all optical wavelengths, high outcoupling eff./detection eff. \textrightarrow{} why using fully tunable FFPIs?: scanning advantageous for inhomo. samples (e.g. membranes, nanoparticles, 2D materials), ease of quickly changing samples\\
Achieved performance of other groups, etc...\\
\end{comment}

\section{Experimental methods}
\label{sec:methods_materials}

Our cavity platform is a home-built design based on a custom-developed nanopositioning platform. The design relies on piezo actuator stacks and electrical DC motors to achieve full tunability of the fiber-based Fabry-Pérot cavity down to cryogenic temperatures. 
%The main working principle is the same for both cavity platforms shown here with slight modifications for each cryostat (see following sections). 
As shown schematically in Fig.~\ref{fig:setup_and_cd_scans}(a), the cavity consists of a macroscopic planar distributed Bragg reflector (DBR) mirror %(Laseroptik GmbH, Hannover) coated onto a superpolished fused silica substrate ($\sigma_{\mathrm{RMS}} < 0.2 \, \mathrm{nm}$) 
which typically carries the sample that shall be investigated. The second mirror consists of a single-mode optical fiber where a concave depression has been machined onto the end facet by CO$_2$-laser machining \cite{hunger_fiber_2010} which then again is coated with a DBR. As the setups presented in this work are designed for the investigation of different samples, we run the experiments with different coatings and laser wavelengths, and thereby also with different design finesses. This influences the required stability and performance of an active feedback since a higher finesse resonance enables a more precise feedback while having the drawback of a smaller stabilization range.
%The expected coating transmission for specific wavelengths and the resulting finesse values can be simulated numerically using a transfer matrix model. 
\begin{comment}
-----------------------------
\textcolor{blue}{(For the flow (closed-cycle) cryostat, we use a fiber F$_A$ (F$_B$) that is coated with a distributed Bragg reflector (DBR) which is designed to have ??? ppm (\SI{950}{ppm}) transmission for the locking wavelength of \SI{640}{nm} (\SI{690}{nm}) and a plane mirror that is designed for a transmission of $??? ppm$ (\SI{750}{ppm}) for the respective wavelength. This corresponds to an expected finesse of $F_{640}\approx ???$ ($F_{690}= 3700$) or a cavity linewidth of $???$ (\SI{93}{pm}) in terms of mirror separation. This linewidth can be used as a benchmark for the desired cavity length stability.) \textrightarrow{} raus?}\\
-----------------------------
\end{comment}

\subsection{Design of the cavity platform}
To achieve highest passive mechanical stability, we have designed a monolithic cavity nanopositioning stage (Fig.~\ref{fig:setup_and_cd_scans}(a),(b)) based on flexure-mechanical elements and piezo actuators for fine tuning that can cover an $(x,y,z)$-volume of $(70\times70\times10)$\,\SI{}{\micro\meter^3} at room temperature that reduces to $\sim(10\times10\times1.5)$\,\SI{}{\micro\meter^3} at \SI{10}{K} (Fig.~\ref{fig:setup_and_cd_scans}(c)). We calibrate the lateral scanning range by cavity transmission images of a planar mirror where a grid pattern with a pitch of \SI{10}{\micro\meter} has been machined by CO$_2$-laser shots. During the cooldown, repeated scanning cavity images reveal a shift of the fiber position to the right due to differential thermal contraction of the piezo actuators and the frame. Also, the scanning range decreases due to the capacity reduction of the piezo stacks.
For coarse tuning over a volume of $(2\times2\times2)$\,\SI{}{mm^3}, the stage incorporates electromotors and a gear work. The central mechanical element is a flexural lever arm to which the cavity fiber is fixed, which itself is glued into a steel needle to increase mechanical stability. The assembly is preloaded by three piezo-electric actuators that allow bending motions to achieve three-dimensional nanopositioning of the fiber tip. The positioning along the cavity axis ($z$) requires particular attention, since it should simultaneously allow one to tune across several longitudinal cavity resonances, i.e. a positioning range of a few micrometers, and to controllably maintain resonance conditions with sub-picometer resolution. The required overall dynamic range of $~10^7$ goes beyond the capability of available voltage sources to drive a single piezo. We thus physically separate the ultra-fine tuning with a thin piezo plate with $\sim 100~$nm expansion range (fine piezo) from the fine tuning with a piezo stack with a few micrometer tuning range (coarse piezo) under cryogenic conditions. Both piezos are separately driven by suitably filtered voltage sources and amplifiers.
To obtain an even wider range of cavity length tunability over a few millimeters, a DC motor, which was slightly modified to work under cryogenic conditions, is used to move the piezo actuators in the longitudinal direction. This wide tuning range is necessary to compensate thermal contraction upon the cooldown of the platform. We minimize the effects of thermal contraction by using titanium to machine the main frame as well as the lever arm, which has a low thermal expansion coefficient (\SI{8.6e-6}{K^{-1}}).
The planar mirror is mounted inside a copper mirror holder which is pressed against the main frame by a spring to ensure mechanical stiffness while keeping the flexibility to move the mirror laterally with two additional DC motors over several millimeters.

\subsection{Active stabilization}
\begin{figure}
	\centering
	\includegraphics[scale=0.35]{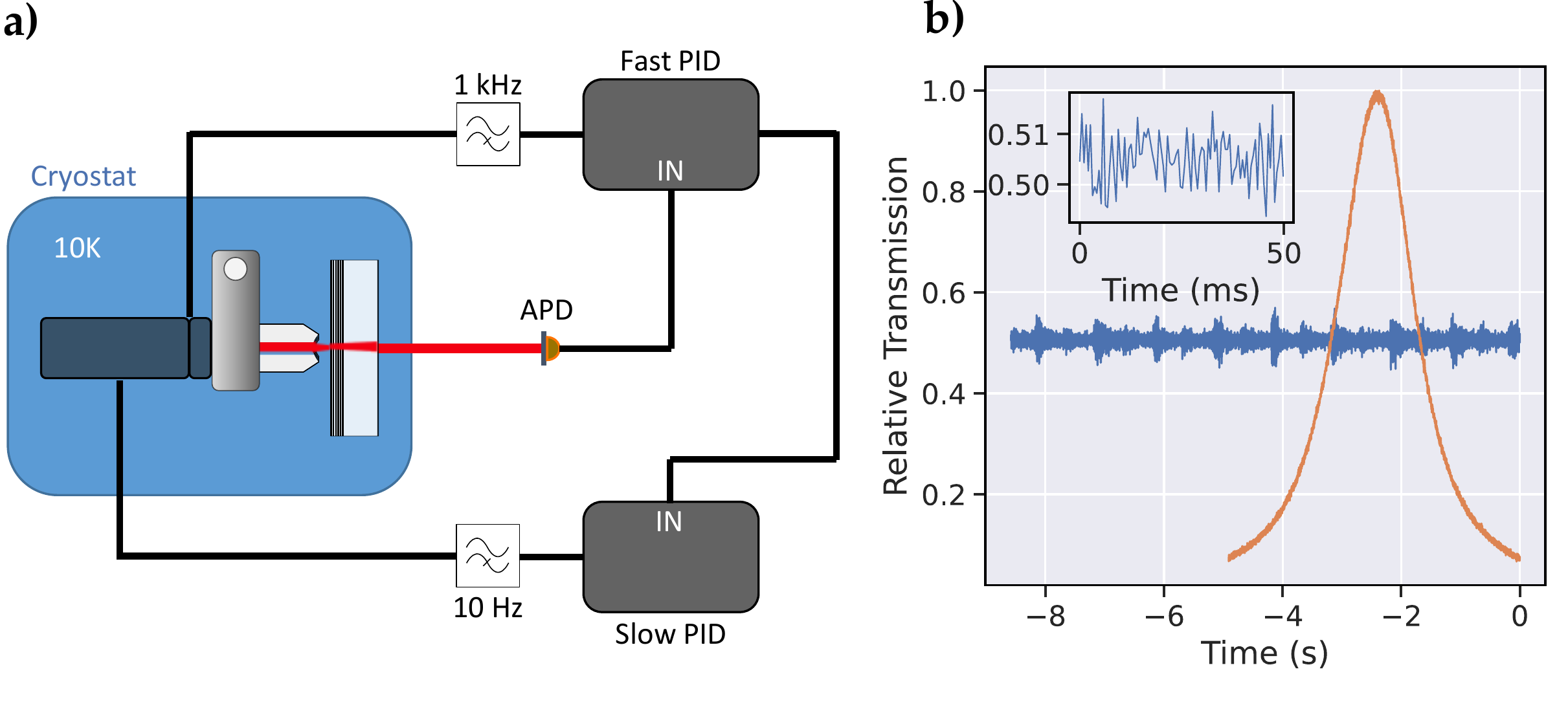}
	\caption{\textbf{(a)} Simplified locking scheme used in this work. The fast PID is realized using a FPGA while the slow lock is software-based in the closed-cycle system. \textbf{(b)} A representative cavity mode with its Lorentzian shape when sweeping the cavity length (horizontal axis in arbitrary units) together with a time trace of the same cavity mode when actively stabilized on the side of the fringe. The inset shows the time trace during the quiet phase of the closed-cycle cryostat.}
	\label{fig:Locking_setup}
\end{figure}

We have investigated different active stabilization schemes, including Pound-Drever-Hall stabilization based on phase modulation as well as side of fringe locking, see Fig.~ \ref{fig:Locking_setup}(a). While the former is insensitive to laser intensity noise and has a larger capture range, it requires high-frequency modulation and a strongly amplified high-bandwidth detector ($1-5~$GHz). Also, it suffers from parasitic Fabry-Perot resonances in the beam path \cite{brachmann_photothermal_2016}. Due to the much lower hardware requirements and the presence of other limiting factors for the lock performance, we thus predominantly use side of fringe stabilization. 
Therefore, we tune the cavity to the slope of a resonance, typically below half of the peak transmission, and generate an error signal with an avalanche photodiode (APD) from the transmitted signal after the plane mirror. For small excursions from the set point, the cavity transmission changes approximately linearly with the cavity length, such that we can both monitor cavity length fluctuations as well as use the signal for active stabilization.
The transmission signal of a Lorentzian cavity resonance when tuning the mirror separation is shown in fig. \ref{fig:Locking_setup}(b), together with an exemplary timetrace of the transmission when actively stabilized to the same resonance while the closed-cycle cryostat is running.\\
The main sources of instability are acoustic and mechanical vibrations and electrical noise - with frequencies ranging from tens of Hz up to tens of kHz - as well as slow drifts, mostly originating from thermal expansion that can have time scales of several hours. We therefore run a combined lock containing two FPGA-based proportional–integral–derivative (PID) controllers (\textit{Red Pitaya STEMlab 125-14}), in the following referred to as fast and slow PID, that can change the cavity length independently through the two piezo actuators controlling the cavity length. In the case of the closed-cycle-system, the slow PID is software-based and runs on a multifunction I/O device (\textit{National Instruments, PCIe-6353}). For the fast PID, we use the cavity transmission signal measured by the APD as a reference signal to suppress the main contributions of noise. The PID output is fed to the locking piezo through a \SI{1}{kHz} low-pass filter. As thermal drifts in the setup occur at a much slower time scale but with larger amplitude, they are counteracted by the slow PID, which works similarly to the fast PID, but using the fast PID output as error signal and being heavily low-pass filtered at \SI{10}{Hz}.
%\textcolor{blue}{Alternative zum obigen Absatz}: For an optimal mechanical stability, a very stiff design is desired in order to minimize the coupling of environmental noise and vibration induced by the cryostat. Therefore, we designed a lever-based, stiff positioning stage that is capable of suppressing high frequency noise, even at cryogenic temperatures. The key aspects here are to implement as few moving parts as possible to reduce possible coupling to noise while still offering full tunability, even at cryogenic temperatures. The fiber mirror F$_A$ is fixed inside a steel needle that improves the mechanical stability of the fiber while providing a robust clamping interface. The needle is held by a movable arm that can be 3-axis-manipulated using piezo stacks. While sacrificing the scanning range that commercially available nanopositioners can offer, the lever-design has a massive advantage in terms of stiffness. To control the mirror separation, a combination of two piezo stacks pushes against the back of the lever, allowing for a scanning range in the \SI{}{\micro\meter}-regime. Furthermore, this allows us to control the cavity length via two independent input channels at the same time, which is a central aspect of our locking scheme and shall be discussed in more detail in section \ref{subsec: noise flow cryo}.

\subsection{Operation in a flow cryostat}
\label{subsec: operation in flow cryo}
One of our fiber cavity platforms is operated inside a customized KONTI micro flow cryostat (CryoVac) and can be cooled down to \SI{3.6}{K}. The cryostat has optical access via two windows such that light can be coupled out of the cavity via the planar mirror. Here, no vibration isolation is incorporated and the cavity platform is placed on posts which are rigidly mounted onto the cryostat cold plate of a helium exchange gas chamber. Cooling via an exchange gas has the advantage of a good thermalization of the sample under investigation and does not require thermal anchoring of the cavity stage to the cold plate. However, it leads to reduced piezo scan ranges since the elongation and force of the piezo stacks reduce with temperature. At the minimum temperature, we reach a longitudinal scan range of about \SI{2}{\micro \meter} i.e. 6 free spectral ranges, and a lateral scanning over an area of \SI{5}{\micro \meter} $\times$ \SI{5}{\micro \meter}. Due to some special treatment of the DC motors, the coarse tuning range of the cavity z-axis and lateral mirror movement doesn't suffer from these low temperatures, so we can maintain the full tunability of the cavity.\\
In this system, a main noise contribution comes from vibrations of the cold plate caused by the liquid helium flow through the heat exchanger beneath the cold plate. By adjusting the impedance of the helium return pipe via a proportional valve, we can reduce the vibrations of the cold plate and thus reach a quiet regime to operate the cavity.

%Piezo: Scan Piezomechanik PSt 150/5x5/20, \SI{1.7}{\micro F}

\subsection{Operation in a closed-cycle cryostat}
\label{subsec: operation in closed cycle}
%Passive mechanical vibration isolation of the cavity platform with simple springs that lead to a strongly damped resonance at a frequency of a few ten Hertz was implemented in the closed-cycle system to suppress the transmission of high-frequency noise to the cavity. Care was taken to use thin and rigidly fixed wiring of the elements on the cavity platform, typically using 0.2~mm thin phosphor bronze wires. Also, a thermalisation link made from a braided mesh of high-conductivity copper filaments connecting the cold plate to the copper mount of the planar cavity mirror was carefully optimized to achieve a minimal temperature gradient and lowest vibration transmission. 
The operation in a closed-cycle cryostat (Montana Cryostation CS1) makes it necessary to implement some adaptions. The vibrations introduced by the cryostat are significantly stronger compared to most flow or bath cryostats and range from frequencies below \SI{100}{Hz} from the compressor itself up to several \SI{10}{kHz}, being composed of electronic and mechanical noise.

The high frequency noise is expected to be more problematic than the lower frequencies, since they cannot be compensated easily via active stabilization. Therefore, the cavity platform is mechanically decoupled from the cold plate using damped springs with a loaded resonance frequency of a few ten Hertz that suppress the transmission of high-frequency noise to the cavity. Care was taken to use thin and rigidly fixed wiring of the elements on the cavity platform, typically using 0.2~mm thin phosphor bronze wires, to avoid mechanical short circuits. %The influence of the springs will be discussed in sec.\ref{sec:results}.\\

In order to cool down the titanium body of the main frame effectively - considering the low thermal conductivity of titanium (\SI{22}{W/mK}) - we thermalize it to the radiation shield of the cryostat using super flexible copper braid stripes (\textit{Copper Braid, SuperFlex \SI{0.2}{mm}}). The flexibility of the braid is crucial as one wants to avoid transmission of vibrations that would counteract the effect of the springs. The braid at the same time acts as the damping system for the springs. \\
In the end, the only part of the cavity that needs to be cooled down to the lowest temperature is the sample, which is usually deposited on the macroscopic planar mirror, while all the other parts of the cavity platform can remain at a higher temperature. %Since fused silica substrates have a low thermal conductivity (\SI{1.38}{W/mK}, taken from Heraeus Quarzglas GmbH), the only way to cool down the sample sufficiently low temperatures is by connecting the mirror to the thermal link of the cryostat and maximizing the area that is connected. 
We therefore use a mirror holder made of copper that is thermally isolated from the main frame by a thermal spacer with minimal contact area. Moreover, a radiation shield made of copper encloses the stage. A thermalization link made from a braided mesh of high-conductivity copper filaments connecting the cold plate to the mirror holder was carefully optimized to achieve a minimal temperature gradient and lowest vibration transmission. The resulting temperature is \SI{10}{K} on the mirror holder, while the base block and the thermal shields show temperatures around \SI{40}{K}, depending on the thermal load by laser radiation or applied current for the motors.
%Since one wants to avoid any uncontrolled motional degree of freedom in the system, especially in longitudinal direction, we firmly pull the mirror holder back to the base block using a steel spring.\\
%For experiments in the closed-cycle cryostat, we additionally make use of the advantages of bringing the fiber tip and the plane mirror into physical contact. While sacrificing some aspects of tunability, e.g. being able to tune the cavity to arbitrary longitudinal cavity mode orders, the mechanical stability is massively enhanced by creating a much more monolithic design.

\section{Results}
\label{sec:results}

For both devices presented here, we measure the mechanical transfer function of the cavity platform to characterize the mechanical resonances that are relevant for the stability of the system. We therefore apply a small sinusoidal voltage to the fine piezo while keeping the cavity length stable with the drift compensation using the coarse piezo. We then measure the transmitted light intensity behind the planar mirror using an APD. The modulation frequency is swept from \SI{100}{Hz} to \SI{100}{kHz} with a logarithmic scale. The APD signal is fed into a lock-in amplifier, where it is mixed with the local oscillator and low-pass filtered to extract the phase and amplitude response of the system. Since the piezo excitation creates oscillations that will disperse across the cavity platform, the resulting transfer function reflects the reaction of the system to mechanical noise in the longitudinal direction, while it does not reveal how strongly vibrations from the cryostat couple to the cavity platform.
%We point out that applying the modulation to the fine piezo does not allow to probe the eigenmodes of the main frame the cavity is mounted at that can be excited through the base plate of the cryostat. Furthermore, lateral eigenmodes of the cavity are also unlikely to be excited in this scheme.\\
We further quantify the mechanical stability of the cavity platform in two ways. First, we apply a drift compensation onto the coarse piezo with a bandwidth of a few Hertz, referred to as passive stability in the following. Second, this is compared to the stability under active stabilization with 1~kHz bandwidth, involving the fine piezo for fast feedback and the coarse piezo for drift compensation. For both techniques we record a time trace of the transmitted intensity through the cavity and investigate its fast Fourier transform (FFT, 100x averaged), yielding a power spectral density (PSD) that we convert into an amplitude spectral density of the cavity length fluctuations by using the first derivative of the Lorentzian cavity resonance at the locking setpoint. 

\begin{comment}
In order to get a meaningful unit for the length jitter, we firstly convert back the ASD values into APD voltages:
\begin{equation}
    U_{APD} = \sqrt{50\Omega \cdot 1mW \cdot 10^{PSD/10}},
\end{equation}
and calculate the voltage rms value from the spectrum,
\begin{equation}
    RMS_{APD} = \sqrt{\sum_{i=0}^{N} U_{APD,i}^2}.
\end{equation}
Due to the side-of-fringe locking scheme, we can transform the voltage jitter into a cavity length jitter. The peak voltage $U_{peak}$, Finesse $F$ and wavelength $\lambda$ of the locking laser fully define the Lorentzian cavity lineshape and thus the 1st derivative at the locking setpoint $K(U_{APD},U_{peak},F,\lambda)$. The rms cavity length jitter is then given by:
\begin{equation}
    RMS = \frac{RMS_{APD}}{K(U_{APD},U_{peak},F,\lambda)}.
\end{equation}
\end{comment}

\subsection{Cavity stability in a flow cryostat}
\label{subsec: noise flow cryo}

\begin{figure}
    \includegraphics[scale=0.35]{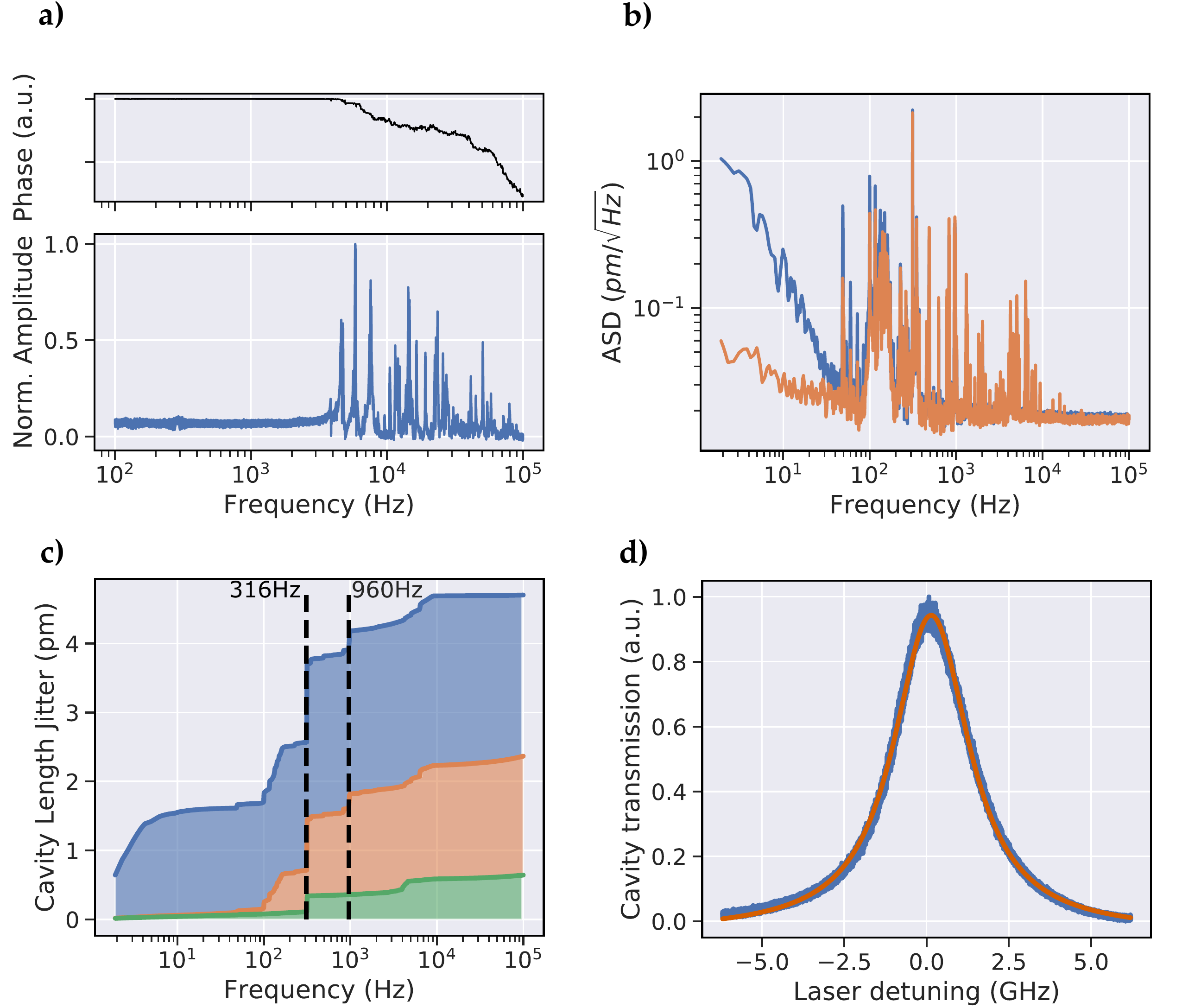}
    \centering
    \caption{Cavity stability at $10K$ in the flow cryostat. \textbf{(a)} Magnitude and phase of the cavity transfer function. \textbf{(b)} Amplitude spectral density of an actively stabilised cavity transmission signal (orange) versus a passively stable cavity (blue). \textbf{(c)} Cumulative cavity length noise obtained from the ASD for the passive cavity (blue), actively stabilized open cavity (orange), and with the fiber in mechanical contact (green). \textbf{(d)} Cavity spectroscopy of the high finesse resonance at \SI{580}{nm} in contact mode.}
    \label{fig:PSD_RMS_flow_cryo}
\end{figure}

Running the flow cryostat at a temperature of \SI{10}{K} in its quiet mode as described above, we now investigate the mechanical stability of the cavity platform. The transfer function of the system, shown in fig.\ref{fig:PSD_RMS_flow_cryo}(a), reveals that the first mechanical resonance occurs at around \SI{4.5}{kHz}, and in the frequency band up to \SI{50}{kHz}, plenty of resonances appear. Achieving such high frequencies of mechanical resonances was a central design principle, since this makes the system inert against the dominant acoustic noise and enables a high bandwidth for active stabilization.%The bandwidth of the PID controller has to be limited to below the first mechanical eigenfrequency because otherwise the feedback signal would drive this resonance due to a phase mismatch.\\

With these findings, we investigated active stabilization and optimized the electrical setup and PID controller settings to achieve a minimal rms value for an actively stabilized cavity. The amplified slow PID output for the drift compensation, which is applied to the coarse piezo, has to be low-pass filtered at a few Hertz to avoid \SI{50}{Hz} noise stemming from the amplifier. We manually optimize the parameters of the fast PID while monitoring the locked cavity length jitter. We find a unity-gain frequency and digital low-pass filtering at \SI{800}{Hz} to give the best value of $\Delta z\approx \SI{2.5}{pm}$ rms. The comparison of the amplitude spectral density (ASD) of the passive (blue) and active (orange) cavity length jitter in Fig.~\ref{fig:PSD_RMS_flow_cryo}(b) reveals a strong effect of the lock for frequencies below \SI{100}{Hz}. For frequencies above \SI{10}{kHz}, we are limited by the noise floor of the measurement electronics and laser noise. To calculate the cumulative rms values in Fig.~\ref{fig:PSD_RMS_flow_cryo}(c), we subtracted an independently measured background spectrum for frequencies above \SI{9}{kHz} since there is no apparent noise contribution from the cavity. The cumulative rms values show that the mechanical resonances around \SI{150}{Hz} and \SI{316}{Hz} are also suppressed by the active feedback. We attribute these resonances and the one at \SI{960}{Hz} to mechanical vibrations of the entire cavity platform and cryostat since they don't appear in the transfer function. The step at \SI{6}{kHz} in the cumulative noise may be attributed to the strongest resonance of the lever arm, which can also be seen in the transfer function in Fig.~\ref{fig:PSD_RMS_flow_cryo}(a).

When studying the cavity without active feedback we find that the passive stability of the setup is very high, reaching a value of $\Delta z\approx \SI{5}{pm}$ rms. Passive stability can be improved by up to an order of magnitude when controllably bringing the fiber in mechanical contact with the planar mirror. Mechanical contact can be monitored by modulating the cavity length and observing the separation of two adjacent cavity resonances when approaching the two mirrors by increasing an offset voltage. As soon as contact is established, the resonances increase their separation. Typically, only one edge or protruding part of the fiber tip will touch, allowing for some bending motion that still allows for resonance tuning, while differential motion is strongly suppressed. Fig.~\ref{fig:PSD_RMS_flow_cryo}(c) shows the cumulative cavity length jitter for such a situation with active stabilization (green) in comparison with measurements without contact.
%  For low frequencies, the lock is limited by the detector noise level, while for high frequencies, the laser intensity noise is the limiting factor. See the supplementary information for a more detailed description of the noise floor.

To illustrate the high mechanical stability when the fiber mirror is brought into contact with its planar opponent, we perform a cavity spectroscopy measurement of a resonance at \SI{580}{nm} which has a high finesse of \SI{16,300}{}. While the cavity resides on the resonance without active feedback, we scan the spectroscopy laser %(DLpro, Toptica GmbH) 
over the full cavity linewidth at a slow rate of \SI{0.5}{Hz} to be susceptible to all noise frequency components. The average of ten scans of the cavity transmission signal is plotted in Fig.~\ref{fig:PSD_RMS_flow_cryo}(d). %The dominant oscillation of the signal is attributed to electrical \SI{50}{Hz} noise stemming from the coarse piezo amplifier since the filtering was still worse when we recorded this measurement.
Fitting a Voigt profile
\begin{comment}
\begin{widetext}
\begin{equation}
T(\nu)= A \mathrm{Re}[w(z)]/(\sigma \sqrt{2\pi}) \quad \text{with} \quad w(z)=exp(-z^2) \textrm{erfc}(-iz), z = \nu -\mu+i\gamma/(\sigma\sqrt{2})
\label{eq:Voigt-profile}
\end{equation}
\end{widetext}
\end{comment}
to the data results in $\Delta z\approx \SI{1.6}{pm}$ deduced from the Gaussian component of the Voigt profile. The data was acquired over a time window of \SI{20}{s}, where mechanical drifts become apparent, leading to a slightly higher cavity length jitter than deduced from the noise spectrum in Fig.~\ref{fig:PSD_RMS_flow_cryo}(b). However, this value confirms the high mechanical stability of the cavity platform over a timespan sufficient for most quantum optics experiments.

% \textcolor{blue}{
% \begin{itemize}
%     \item (a): transfer function at 10K; sweeping sine-frequency on fine z-piezo in logarithmic sweep from 100Hz to 100kHz within 16s; sit passively on sof with battery connected to coarse z-piezo; interpretation: no resonance below 4.5kHz visible hence 310Hz resonance must come from cavity stage assembly or cryo-mechanics; lots of resonances in band between 4.5kHz and 50kHz which we attribute to the lever arm and piezo mechanical modes.
%     \item (b): locked cavity; detector+osci noise subtracted; 100x averaged; laser noise subtraction couldn't really remove tail at high frequencies; 50Hz peak from coarse z-piezo amplifiers; 310Hz peak from whole stage assembly since it gets worse at cold temperatures (cryo-plate resonance, stiffer assembly); 6.5kHz resonance from lever arm + piezo mechanics which gets stiffer at low temperatures 
%     \item (c): active-passive comparison at 10K; passive = PID switched off; PID is reducing laser noise for $f<50$Hz and 50Hz peak; no reduction of 310Hz peak and 6.5kHz resonance (locking bandwidth: 800Hz)
% \end{itemize}}

\subsection{Cavity stability in a closed-cycle cryostat}
\label{subsec: noise closed-cycle cryo}
\begin{figure}
    \includegraphics[scale=0.35]{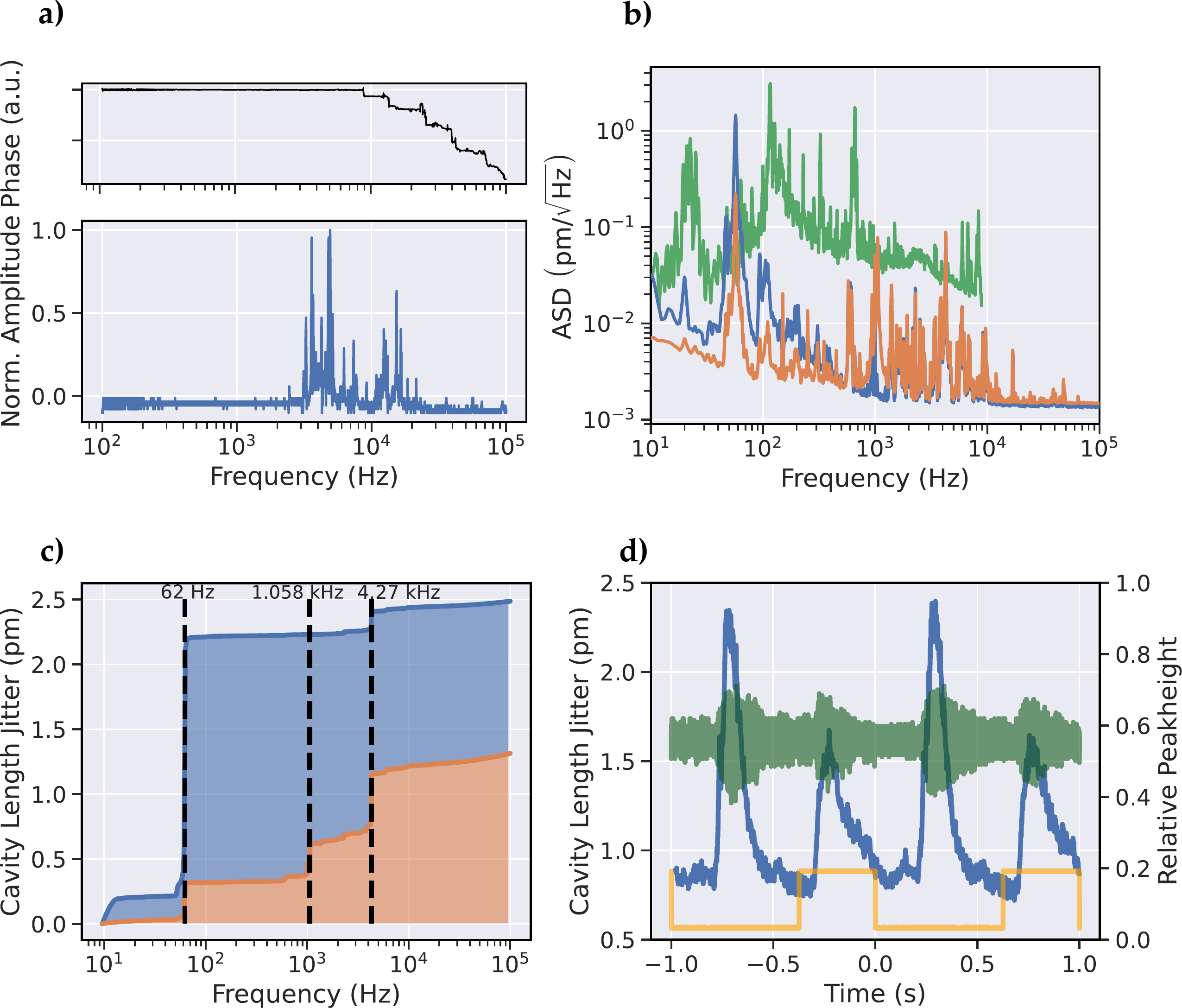}
    \centering
    \caption{Cavity stability at \SI{10}{K} in the closed-cycle cryostat. \textbf{(a)} Mechanical transfer function of the cavity. \textbf{(b)} ASD of the cavity length fluctuations without (blue, \SI{2.4}{pm\,rms}), with active stabilization (red, \SI{1.2}{pm\,rms}) and without contact of the fiber and plane mirror (green, \SI{15}{pm\,rms}). \textbf{(c)} Cumulative cavity length noise for the respective measurements in (b). \textbf{(d)} Time-resolved cavity noise over two cold head cycles. Yellow: trigger output of the cryostat. The cavity transmission (green) is shown together with the cavity length fluctuation (blue).}
    %to do: normalize transfer function amplitude to 1
    %evtl. motion ASD anstatt PSD
    \label{fig:PSD_RMS_closed_cycle}
\end{figure}

We perform analogous measurements for the cavity setup in the closed-cycle cryostat.
The transfer function (Fig.~\ref{fig:PSD_RMS_closed_cycle}(a)) does not show any resonances below \SI{4}{kHz}, which is in accordance to the measurements on the flow cryostat setup and verifies the reproducibility of the design. The highest eigenfrequencies appear around \SI{20}{kHz}. %Note that this measurement excludes vibration eigenmodes that originate from the bottom part of the cavity stage which are coupled to the top part via the springs, as the source of vibrations is the fine piezo located on the top part.\\ 

The ASD spectra of our best performing configurations with the cavity mirrors in mechanical contact, and with optimized clamping of cables, pre-stress on screws, electric filtering etc., are shown in Figure~\ref{fig:PSD_RMS_closed_cycle}(b). Both the passive (blue) and actively stabilized (orange) configuration show a significant improvement in noise compared to the actively stabilized operation without mechanical contact of the mirrors (green).

Similarly to the flow cryostat setup, the fast PID output is low-pass filtered at \SI{1}{kHz}, and the low noise level above \SI{1}{kHz} shows the high passive stability of the system. We observe an increase in noise around \SI{1}{kHz} depending on the PID gain, which is due to the servo bump of the feedback loop and limits the gain of the fast PID.\\

In the cumulative depiction, see Figure~\ref{fig:PSD_RMS_closed_cycle}(c), three main frequencies stand out: the largest fraction of noise in the passive mode at \SI{62}{Hz} originates from the compressor that pumps the helium into the cold head, the servo bump at \SI{1.058}{kHz} and a third contribution at \SI{4.27}{kHz} which we attribute to the dominant resonance of the lever arm. Note that the servo bump is absent for the passive configuration. The biggest improvement of the stabilization is a significant decrease of noise introduced by the compressor, showing an almost 10-fold lower rms noise up to that frequency. By integrating over the whole frequency range, we extract a passive stability over the whole cryostat cycle of \SI{2.4}{pm\,rms} which improves to \SI{1.2}{pm\,rms} when actively stabilized, in contrast to \SI{15}{pm\,rms} without contact.\\

By synchronizing the measurement to the trigger output of the cryostat, we can examine the cavity noise temporally resolved throughout the cold-head cycle. When considering only the quiet phase of the cycle, which spans over several \SI{100}{ms} and is therefore long enough for most spectroscopy experiments, the measurement reveals a cavity length jitter as low as \SI{0.8}{pm} rms, see Fig.~\ref{fig:PSD_RMS_closed_cycle}(d), corresponding to a noise of $<\SI{1}{\%}$ of the cavity linewidth at a finesse of 3700 as used for the stabilisation. %To our knowledge, this value represents a record stability of a fiber-based Fabry-Pérot microcavity inside a closed-cycle cryostat.
The cavity parameters of both setups are summarized in table \ref{tab:cavity_parameters}.

\begin{table}[h]
    \centering
    \begin{tabular}{|c c c|}
        \hline
        \textbf{Parameter} & \textbf{Flow cryostat} & \textbf{Closed-cycle cryostat}\\[1ex]
        \hline
        Lock wavelength & \SI{640}{nm} & \SI{690}{nm} \\
        Lock-Finesse &  1200 & 3700 \\
        Lock-FWHM & \SI{260}{pm} & \SI{90}{pm} \\
        Max. Finesse & 18000 & 37000 \\
        FWHM & \SI{15}{pm} & \SI{8.6}{pm} \\
        $\Delta z_{open,passive}(10K)$ & \SI{5}{pm} & \SI{15}{pm} \\
        $\Delta z_{open,active}(10K)$ & \SI{2.5}{pm} & \SI{15}{pm} \\
        $\Delta z_{contact,passive}(10K)$ & \SI{0.8}{pm} & \SI{2.4}{pm} \\
        $\Delta z_{contact,active}(10K)$ & \SI{0.7}{pm} & \SI{1.2}{pm} (\SI{0.8}{pm}) \\
        \hline
    \end{tabular}
    \caption{Summary of the cavity parameters for the platform installed in a flow cryostat and a closed-cycle cryostat.}
    \label{tab:cavity_parameters}
\end{table}

\section{Conclusion}
\label{sec:conclusion}

In this work, we presented a home-built open-access fiber-based Fabry-Pérot microcavity platform which achieves highest mechanical stability of better than \SI{1}{pm}, both in a closed-cycle cryostat and in a flow cryostat. The basis therefore is a mechanical design of high intrinsic stiffness, the incorporation of active feedback stabilization, and by bringing the fiber into mechanical contact with the planar mirror. We have demonstrated the reproducibility of our design principles by successfully operating two cavity stages inside two different cryostat types.
% This platform can be actively stabilized down to a cavity length jitter of \SI{2.5}{pm} rms without mechanical contact. % By doing so, we expect a large improvement in stability also for the flow cryostat stage down to the sub-pm regime. Thus, we would be capable of operating cavities with finesses of up to \SI{100,000}{} under cryogenic conditions without much detoriation of the Purcell-effect by vibrational linebroadening.\\

The passive stability of a few \SI{}{pm} of both setups is already good enough to resign from active stabilization, which requires a second locking laser and a double-resonance condition with the spectroscopy laser. The vibration broadening of the cavity linewidth for a finesse of \SI{20,000}{} would only reduce the Purcell-factor by less than \SI{5}{\percent} when both mirrors are in contact, and about \SI{10}{\percent} without mechanical contact in the flow cryostat, respectively (see Fig.~\ref{fig:Purcell_vs_RMS}). %Thus we can perform cavity-enhanced spectroscopy experiments in a pulsed mode and use the spectroscopy laser for the slow drift compensation. Using this scheme in the flow cryostat we gain back the full longitudinal tunability and can operate the cavity at any resonance. For certain quantum emitters like Europium ions this is advantageous since a double-resonance of two atomic transitions at a certain cavity length can be chosen, leading to a twofold Purcell-enhancement.\\

Both designs combine excellent mechanical stability with full three-axis coarse and fine tunability at cryogenic temperatures. Furthermore, the cavity stage facilitates the fast exchange of the planar mirror carrying the sample under investigation. Therefore, this platform is particularly useful for the investigation of many different, spatially and spectrally inhomogeneous samples, % like nanoparticles, (molecular) thin-films, membranes, TMDs or semiconductors
where lateral scanning is needed to find suitable emitters, and longitudinal tuning is required to target the desired transition. This holds great prospects to enable fundamental studies in the field of solid-state quantum optics and advanced applications in quantum technologies.\\

\section*{Author Declarations}
\subsection*{Conflict of Interest}
The authors have no conflicts to disclose.
\subsection*{Author Contributions}
\textbf{Maximilian Pallmann:} Conceptualization; Data curation; Formal analysis; Investigation; Methodology; Software; Validation; Visualization; Writing - original draft; Writing - review \& editing. \textbf{Timon Eichhorn:} Conceptualization; Data curation; Formal analysis; Investigation; Methodology; Software; Validation; Visualization; Writing - original draft; Writing - review \& editing. \textbf{Julia Benedikter:} Conceptualization; Methodology; Software; Writing - review \& editing. \textbf{Bernardo Casabone:} Conceptualization; Methodology; Writing - review \& editing. \textbf{Thomas Hümmer:} Conceptualization; Methodology; Software; Writing - review \& editing. \textbf{David Hunger:} Conceptualization; Funding acquisition;  Methodology; Project administration; Resources; Supervision; Validation; Writing - review \& editing.

\section*{Data Availability Statement}
The data that support the findings of this study are available from the corresponding author upon reasonable request.

\section*{Acknowledgements}
 We thank the team of Qlibri GmbH for fruitful discussions, which commercializes a modified version of the reported cavity platform. We also thank Chetan Deshmukh, Eduardo Beattie, Hugues de Riedmatten and Khaled Karrei for many fruitful discussions. We acknowledge support from Leonhard Neuhaus to adapt the Pyrpl software package to our needs. This work has been financially supported by the European Union Quantum Flagship initiative under grant agreement No. 820391 (SQUARE), the BMBF projects q.link.x (grant agreement No. 16KIS0879), QR.X (grant agreement No. 16KIS004), and SPINNING (grant agreement No. 13N16211), the Karlsruhe School of Optics and Photonics (KSOP), and the Max-Planck School of Photonics.

\bibliography{ref,referencesDavid}
%\bibliography{referencesDavid}

\end{document}